\documentclass[11pt]{article}

\usepackage[left=2cm, right=2cm, top=2.5cm, bottom=2.5cm]{geometry}
\usepackage[x11names]{xcolor}
\usepackage{fancyhdr, amssymb, cancel, amsmath, graphicx, pgfplots, tikz, bm}

\newcommand{\stylecolor}{black}

\usepackage[colorlinks=true, urlcolor=violet, linkcolor=blue, citecolor=red, hyperindex=true, linktocpage=true]{hyperref}

\newcommand{\ccore}{{\chi_0}}
\newcommand{\rstar}{\bar r}
\newcommand{\ii}{\mathrm{i}}
\newcommand{\ee}{\mathrm{e}}

\usepackage[explicit]{titlesec}

\newcommand*\sectionlabel{}
\titleformat{\section}
  {\gdef\sectionlabel{}
   \Large\bfseries\scshape}
  {\gdef\sectionlabel{\thesection. }}{0pt}
  {\begin{tikzpicture}[remember picture,overlay]
	\draw (-0.2, 0) node[right] {\textsf{\sectionlabel#1}};
	\draw[thick] (0, -0.4) -- (\textwidth, -0.4);
       \end{tikzpicture}
  }
\titlespacing*{\section}{0pt}{15pt}{20pt}

\newcommand*\subsectionlabel{}
\titleformat{\subsection}
  {\gdef\subsectionlabel{}
   \large\bfseries\scshape}
  {\gdef\subsectionlabel{\thesubsection.\ \  }}{0pt}
  {\begin{tikzpicture}[remember picture,overlay]
    	\draw (-0.15, 0) node[right] {\textsf{\subsectionlabel#1}};
       \end{tikzpicture}
  }
\titlespacing*{\subsection}{0pt}{10pt}{10pt}

\newcommand*\subsubsectionlabel{}
\titleformat{\subsubsection}
  {\gdef\subsubsectionlabel{}
   \bfseries\scshape}
  {\gdef\subsubsectionlabel{\thesubsubsection.\ \  }}{0pt}
  {\begin{tikzpicture}[remember picture,overlay]
    	\draw (-0.15, 0) node[right] {\textsf{\subsubsectionlabel#1}};
       \end{tikzpicture}
  }
\titlespacing*{\subsubsection}{0pt}{7pt}{7pt}

\pgfplotsset{every axis legend/.append style={at={(1.02,1)},anchor=north west}}

\newcommand{\titletext}{Sound-induced vortex interactions in a zero temperature two-dimensional superfluid}

\begin{document}

\pagestyle{fancy}
\renewcommand{\headrulewidth}{0pt}
\fancyhead{}

\fancyfoot{}
\fancyfoot[C] {\textsf{\textbf{\thepage}}}

\begin{equation*}
\begin{tikzpicture}
\draw (0.5\textwidth, -3) node[text width = \textwidth] {{\huge \begin{center} \color{\stylecolor} \textsf{\textbf{\titletext}} \end{center}}};
\end{tikzpicture}
\end{equation*}
\begin{equation*}
\begin{tikzpicture}
\draw (0.5\textwidth, 0.1) node[text width=\textwidth] {\large \color{black} \textsf{Andrew Lucas and Piotr Sur\'owka}};
\draw (0.5\textwidth, -0.5) node[text width=\textwidth] {\small \textsf{Department of Physics, Harvard University, Cambridge, MA, USA 02138}};
\end{tikzpicture}
\end{equation*}
\begin{equation*}
\begin{tikzpicture}
\draw (0.5\textwidth, -6) node[below, text width=0.8\textwidth] {\small We present a systematic derivation of the effective action for interacting vortices in a non-relativistic two-dimensional superfluid described by the Gross-Pitaevskii equation by integrating out longitudinal fluctuations of the order parameter.   There are no logarithmically divergent coefficients in the equations of motion.    Our analysis is valid in a dilute limit of vortices where the intervortex spacing is large compared to the core size, and where number fluctuations of atoms in vortex cores are suppressed.    We analyze sound-induced corrections to the dynamics of a vortex-antivortex pair and show that there is no instability to annihilation, suggesting that sound-mediated interactions are not strong enough to ruin an inverse energy cascade in two-dimensional zero-temperature superfluid turbulence.
};
\end{tikzpicture}
\end{equation*}
\begin{equation*}
\begin{tikzpicture}
\draw (0, -13.1) node[right, text width=0.5\paperwidth] {\texttt{lucas@fas.harvard.edu \\ surowka@physics.harvard.edu}};
\draw (\textwidth, -13.1) node[left] {\textsf{\today}};
\end{tikzpicture}
\end{equation*}

\tableofcontents

\section{Introduction}
Recent experiments \cite{becexp2, becexp} on two-dimensional turbulent superfluids have increased the importance of resolving fundamental theoretical questions about the nature of superfluid turbulence in two dimensions.   One of the most basic outstanding questions in the field concerns whether energy is convected towards large length scales or small length scales.   The former is characteristic of the enstrophy-conserving inverse energy cascade of normal two-dimensional fluids \cite{kraichnan}.  Some numerical simulations have suggested that this is indeed the case \cite{reeves, billam, simula}, while others \cite{numasoto, chesler} have emphasized the role of vortex annihilation in driving energy to small length scales.

Our goal in this paper is to help settle this question in a non-relativistic superfluid at zero temperature.  Crucial to this task is a proper understanding of the effective dynamics of vortices in the superfluid.  The effective equation of motion of a superfluid vortex is a rather controversial question with a long history in the literature \cite{hallvinen, iordanskii, popov, halperin1, halperin2, chandler, sonin, wexler, thouless, thompson}, even in two spatial dimensions.  Furthermore, almost all of this previous work focuses on single vortex dynamics, yet superfluid vortices have long range interactions, suggesting that an understanding of single vortex dynamics would nevertheless not suffice for understanding multi-vortex dynamics, beyond-leading order.

In this paper, we present a systematic calculation of the effective action of $N> 1$ superfluid vortices, assuming that the underlying continuum action is the Gross-Pitaevskii (GP) action \cite{gross, pitaevskii}.   This non-relativistic action is used in nearly all simulations of superfluid turbulence \cite{reeves, billam, simula, numasoto}, and so serves as a natural choice.  Our perturbative parameter is the ratio of the vortex core size to intervortex spacing, $\xi/\rstar$; our calculation is valid at next-to-leading order ($\mathcal{O}(\xi^2/\rstar^2)$), and so takes into account the leading-order dressing of superfluid vortices by sound.

Similar papers have recently described effective vortex-sound interactions by effective action techniques \cite{sonwingate, rattazi, hui} in a three dimensional normal fluid \cite{nicolis}, and more recently in a three dimensional superfluid \cite{gubser, nicolis2}.   We should note that the extension of our calculation to three dimensions is complicated by Kelvin waves -- normal mode excitations of the stringy vortices \cite{vinen, kozik1, kozik2}.   Effective action techniques have also been used to study vortices in two-dimensional superconductors \cite{nikolic} and in p-wave superfluids \cite{ariad}.

Our paper is organized as follows.  Section \ref{gpsection} reviews GP theory and provides set-up and notation for our computation.  Section \ref{actionsection} outlines the computation of the effective action at next-to-leading order and summarizes our results.   An interesting observation that we find is that the notion of ``vortex mass" becomes ill-defined, with kinetic terms coupling the velocities of distinct vortices.  Furthermore, no coefficients in the equations of motion (on-shell) have any logarithmic divergences in the microscopic core size or a macroscopic ``box size".    Section \ref{smokesection} describes the dynamics of a vortex-antivortex pair, which can be found exactly at next-to-leading order.   Our main result is that this pair will not annihilate.  This suggests that the inverse cascade description of turbulence is appropriate at zero temperature.   Appendices provide pedagogical computations, as well as the details of our calculation.

\section{Gross-Pitaevskii Equation}\label{gpsection}
We begin with the GP action, which is the simplest theory of a non-relativistic superfluid phase, at zero temperature (in units with $\hbar=1$): \begin{equation}
S = \int \mathrm{d}^2\bm x\mathrm{d}t\; \left[\bar{\psi} \left(\ii \partial_t + \frac{\nabla^2}{2m} + \mu \right)\psi - \frac{\lambda}{2}\left(\bar{\psi}\psi\right)^2\right].
\end{equation}
The equations of motion from this action are \begin{equation}
\ii\partial_t \psi = -\frac{1}{2m}\nabla^2\psi - \mu \psi + \lambda |\psi|^2 \psi.  \label{gpe}
\end{equation}The vacuum of this theory is described by the superfluid phase so long as $\mu>0$.  In this phase, there is a non-vanishing background superfluid density of \begin{equation}
\rho_0 \equiv \left\langle |\psi|^2\right\rangle_{\text{vacuum}} = \frac{\mu}{\lambda}.
\end{equation}Note that the phase of $\psi$ is undetermined.   As we will see, it is helpful to make a change of variable to \begin{equation}
\psi = \sqrt{\rho_0} \ee^{\chi + \ii \theta},\;\;\; \bar{\psi} = \sqrt{\rho_0} \ee^{\chi - \ii\theta},
\end{equation}where $\chi$ and $\theta$ are real-valued fields, which keep track of density and phase fluctuations respectively.   Note that $\theta \equiv \theta+2\pi$ describe the same physics and are thus equivalent.

It is helpful to express the Gross-Pitaevskii action and equation in terms of $\chi$ and $\theta$.   The action can be worked out straightforwardly: up to total derivatives, the answer is
\begin{equation}
S = -\frac{\mu}{\lambda}  \int \mathrm{d}^2 \bm x\mathrm{d}t \,
\ee^{2\chi} \left[ \partial_t \theta + \frac{(\nabla\theta)^2}{2m}+\frac{(\nabla\chi)^2}{2m}
+ \mu \left(\frac{\ee^{2\chi}}{2}-1 \right) \right] \label{eq5}
\end{equation}
The equations of motion follow from this action: \begin{subequations}\begin{align}
\frac{\delta S}{\delta \chi} &\equiv J_\chi = -\frac{2\mu}{\lambda} \ee^{2\chi}
\left[\partial_t\theta + \frac{(\nabla\theta)^2}{2m} + mP\right]= 0,\\
\frac{\delta S}{\delta \theta} &\equiv J_\theta = \frac{2\mu}{\lambda}\ee^{2\chi}\left[\partial_t \chi + \frac{\nabla\theta}{m}\cdot\nabla\chi + \frac{\nabla^2\theta}{2m}\right] = 0.   \label{eqjtheta}
\end{align}\end{subequations}where we introduce the ``pressure",\begin{equation}
P= \frac{1}{m}\left[\mu \left(\ee^{2\chi}-1\right) - \frac{(\nabla \chi)^2 + \nabla^2\chi}{2m} \right]
\end{equation}We will often also define the superfluid velocity, \begin{equation}
\bm v = \frac{\nabla\theta}{m}.
\end{equation}
In terms of the superfluid density,\begin{equation}
\rho = \rho_0\ee^{2\chi},
\end{equation} we can write the above equations in ``hydrodynamic" form: \begin{subequations}\begin{align}
\partial_t \rho + \nabla\cdot (\rho \bm v) &= 0, \\
\partial_t \bm v + (\bm v \cdot \nabla) \bm v &= -\nabla P.
\end{align}\end{subequations}
This suggests that, in a limit where core physics can be neglected, solutions of classical hydrodynamics should be a good approximation to the dynamics.   We will see that this is indeed the case, in this section.

The fact that $\theta$ is identified with itself upon integer multiplies of $2\pi$ allows for non-trivial topological solutions  to Eq. (\ref{gpe}) called vortices.    If we place a vortex of winding number $\Gamma \in \mathbb{Z}$ at the origin $x=y=0$, we can find a stationary solution to Eq. (\ref{gpe}), if we make the ansatz \begin{equation}
\theta = \Gamma \arctan \frac{y}{x}, \;\;\;\;\; \chi = \ccore\left(\sqrt{x^2+y^2}\right),
\end{equation}where $\Gamma = \pm 1$ denotes the orientation of the vortex.   $\ccore$ is the solution to\footnote{An identical ansatz can be made for $|\Gamma|>1$ vortices, although the function $\ccore$ changes.} \begin{equation}
\ccore^{\prime\prime} + \chi^{\prime2}_{0} + \frac{\chi^\prime_{0}}{r} - \frac{1}{r^2} = \frac{\ee^{2\ccore}-1}{\xi^2}, \label{chicore}
\end{equation}with boundary conditions that $\ccore(\infty) = 0$, $\ccore(0)=-\infty$.   We will not consider vortices with $|\Gamma| >1$ as these vortices are unstable and will break up into winding number $\pm 1$ vortices rapidly during the evolution of the superfluid condensate (see, e.g., \cite{chesler}).     Defining a ``healing length", \begin{equation}
\xi^2 \equiv \frac{1}{2m\mu},
\end{equation}which is the only length scale in Eq. (\ref{chicore}), we can easily determine the asymptotic behavior of $\ccore$: \begin{equation}
\ccore(r) \approx -\frac{\xi^2}{r^2}\;\;\;(r\gg \xi), \;\;\;\;\;\; \ccore(r) \approx \log\frac{r}{\xi},\;\;\;(r\ll \xi).
\end{equation}

\subsection{The Point Vortex Ansatz}
We are interested in computing the effective action of $N$ vortices within this framework.   Let us suppose that we are in a \emph{dilute limit} where the density of vortices is very small.   We will quantify this limit shortly.   In this limit, we expect that there is an approximate solution to Eq. (\ref{gpe}) of the form \cite{bradley},
\begin{subequations} \label{eq13}\begin{align}
\chi_{\mathrm{PV}}(\bm x,t) &= \sum_{n=1}^N \ccore(\bm x-\bm X_n(t)) \equiv \sum_{n=1}^N \chi_n, \\
 \theta_{\mathrm{PV}}(\bm x,t) &= \sum_{n=1}^N \Gamma_n \arctan \frac{y-Y_n(t)}{x-X_n(t)} \equiv \sum_{n=1}^N \Gamma_n \theta_n.
\end{align}\end{subequations}
In this paper, we will use $n$ to denote individual vortices, and $\bm X_n(t)$ to denote their trajectories.    Sometimes we will also use index notation for $\bm X_n$:  $X_n^i$, where $i$ denotes vector indices.  On this ansatz, $\bm v$ may be written as a sum over contributions $\bm v_n$ from each vortex.      As we will see, this ansatz is indeed an asymptotically good approximation in the dilute limit, if $X_n^i$ obeys point-vortex dynamics \cite{aref}:
\begin{equation}
\dot{X}^i_n = -\epsilon^{ij} \sum_{m\ne n} \frac{\Gamma_m}{m} \frac{X_n^j - X_m^j}{|\bm X_n - \bm X_m|^2} = \sum_{m\ne n} v^i_m(X_n) \equiv \sum_{m\ne n} V^i_{m,n} \equiv U^i_n,  \label{uvdefs}
\end{equation}
where $\bm v_n \equiv \nabla \theta_n /m$.   We have also defined $\bm U_n(\bm X_m)$ to be the superfluid velocity through core $n$ -- note that it depends on the position of every single vortex.   We are also using the two-dimensional Levi-Civita symbol $\epsilon^{xy} = -\epsilon^{yx}=1$, $\epsilon^{xx}=\epsilon^{yy} = 0$.

In this paper, we will denote the typical distance from any vortex to its nearest neighbor as $\rstar$.   The typical velocity scale is thus $1/m\rstar$, and the typical time scale is $m\rstar^2$.

Let us now quantify the dilute limit, where $|\psi^2| = \rho_0 \mathrm{e}^{2\chi} \approx \rho_0$, or $\chi \approx 0$.  Assuming that we have a uniformly random distribution of vortices with density $1/\rstar^2$, we can replace \begin{equation}
\sum_{m\ne n} F(\bm X_m - \bm X_n) \approx \frac{1}{\rstar^2} \int\limits_{\rstar}^{r_{\mathrm{max}}}\mathrm{d}^2\bm x \; F(r)  \label{averageeq}
\end{equation} for an arbitrary function $F$.  Here $r_{\mathrm{max}}$ is a scale denoting the size of the cluster of vortices, e.g., proportional to the largest distance between any two vortices.   Now, let us apply this to the field $\chi$.  We find that the dilute limit corresponds to \begin{equation}
\sum_n \chi_n \approx 0 \approx \frac{1}{\rstar^2} \int\limits_{\rstar}^{r_{\mathrm{max}}} \mathrm{d}^2\bm x \; \frac{\xi^2}{r^2} \sim \frac{\xi^2}{\rstar^2} \log \frac{r_{\mathrm{max}}}{\rstar},
\end{equation}
which gives us the dilute limit corresponding to
\begin{equation}
\frac{\rstar^2}{\xi^2} \gg \log \frac{r_{\mathrm{max}}}{\rstar}.   \label{diluteeq}
\end{equation}

We will always assume in this paper that we are in a dilute limit.   Within pure point vortex dynamics, there may be close passes between a vortex-antivortex pair.    If such a close pass occurs for a vortex pair, it may mean that the computation does not hold for that pair.


\subsection{The Action of Point Vortex Dynamics}


It is appreciated (see, e.g., \cite{fetter}) that point vortex dynamics is an asymptotically good solution to the GPE.  In Appendix \ref{checkapp}, we provide a careful check that this is indeed true; parts of the calculation are also helpful for our main computation.    The leading order action for vortices is the $\chi$-dependent contribution, \begin{equation}
S_{\mathrm{PV}} \approx -\frac{\mu}{\lambda} \int \mathrm{d}^2\bm x\mathrm{d}t \; \left[\partial_t \theta + \frac{(\nabla \theta)^2}{2m}\right].
\end{equation}As we will now show, this precisely reproduces the action for point vortex dynamics.

We begin by studying the $(\nabla\theta)^2$ integral.   This divergent integral must be regulated by an IR cutoff, and can be done by dimensional analysis.   This integral is performed in Appendix \ref{dimreg} (see Eq. (\ref{i21})), and the answer is \begin{equation}
S_{\mathrm{PV,pot}} = \pi\rho_0\int \mathrm{d}t \; \sum_{m\ne n} \frac{\Gamma_m\Gamma_n}{m}\log\frac{|\bm X_m-\bm X_n|}{L}.
\end{equation}
There are two ways to obtain the kinetic term in the action.  The first is simply to pick the answer which ensures that point vortex dynamics are the equations of motion.   Alternatively, one can use a different regulatory scheme\footnote{Although this integral looks badly divergent in the IR, it is straightforward to regulate by restricting the spatial integral to a box of size $L$, and then taking $L\rightarrow \infty$ at the end of the calculation.   Ignoring constants and total derivative terms $S_{\mathrm{PV,kin}}$ is the leading order answer in $L$.   The reason that the standard dimensional regularization technique employed in the appendix will not work for this integral is that the integrand (the Lagrangian) is not translation invariant.} and one finds\begin{equation}
S_{\mathrm{PV,kin}} = \pi \rho_0 \int \mathrm{d}t \; \sum_n \Gamma_n \frac{\epsilon_{ij}\dot{X}^i_n X^j_n}{2}.
\end{equation}Thus we obtain \begin{equation}
S_{\mathrm{PV}} = \pi\rho_0 \int \mathrm{d}t \; \left[\sum_n \Gamma_n \frac{\epsilon_{ij}\dot{X}^i_n X^j_n}{2} + \sum_{m\ne n} \frac{\Gamma_m\Gamma_n}{m} \log\frac{|\bm X_m-\bm X_n|}{L} \right] \label{spv}
\end{equation}

Noether's Theorem may be straightforwardly used to generate conserved quantities.   Some important conserved quantities in point vortex dynamics follow from time translation invariance:  the energy $E$ is given by  \begin{equation}
E = -\pi\rho_0 \sum_{m\ne n} \frac{\Gamma_m\Gamma_n}{m} \log \frac{|\bm X_m - \bm X_n|}{L}.
\end{equation}Translation invariance under $\bm X_m \rightarrow \bm X_m + \bm a$ (for all $m$) gives an analog of momentum conservation:\footnote{The action in this case is invariant, although the Lagrangian is not.} \begin{equation}
 P^i = \epsilon^{ij}\pi \rho_0 \sum_m \Gamma_m X^j_m.  \label{momcons}
\end{equation}

One may notice that the form of Eq. (\ref{spv}) is very similar to the electrodynamics of massless charged particles of charge $\Gamma$.   This is a manifestation of the particle-vortex duality \cite{popov, fisher} between the low energy effective theory of a superfluid and relativistic electrodynamics with charged scalars.   It will not remain when we compute corrections to the action below.

Finally, we point out that our answers may seem somewhat surprising -- although the GP action was manifestly Galilean invariant, point vortex dynamics is not so.   This is a consequence of the fact that the placement of a vortex picks out a preferred rest frame.   We discuss in Appendix \ref{galilean} how to restore Galilean invariance manifestly in point vortex dynamics.

\section{The Effective Action}\label{actionsection}
Now, we are ready to compute the effective action for vortices.   To do this, we take our ansatz for $\chi$ and $\theta$ on point vortex dynamics, as defined in the previous section, and integrate out fluctuations in $\chi$ and $\theta$ in a path integral formalism.   More precisely:  let \begin{subequations}\begin{align}
\chi &= \chi_{\mathrm{PV}} + \delta \chi, \\
\theta &= \theta_{\mathrm{PV}} + \delta \theta.
\end{align}\end{subequations}The $\delta \theta$ contribution does not contain any singularities (vortices) -- i.e. it is a smooth single valued function everywhere in space and time.   Time-dependent solutions of the GP equation may be found by finding saddle point solutions to a path integral \cite{peskin} \begin{equation}
Z = \int \mathrm{D}\chi\; \mathrm{D}\theta\; \ee^{\ii S[\chi,\theta]}.
\end{equation}subject to appropriate boundary conditions at the initial time (since the GP equation is first order).  Effective action techniques use the fact that as a path integral, we may selectively integrate over some of the $\chi$ and $\theta$ modes \emph{before} finding the minima of $S$.   In this paper, we will integrate over the $\delta \chi$ and $\delta \theta$ modes, while leaving the modes $\bm X_n(t)$ free to fluctuate.  It is only over these $\bm X_n$ modes that we will find the saddle point of $Z$.

We will perform this calculation at one-loop order -- i.e., we will approximate the true path integral over $\delta \theta$ and $\delta \chi$ by a Gaussian path integral, as follows.   We perform a Taylor expansion of $S[\chi,\theta]$: \begin{equation}
S[\chi,\theta] = S_{\mathrm{PV}}[\bm X_n] + \int \mathrm{d}^2\bm x \mathrm{d}t\left[\left(\begin{array}{cc} J_\chi &\ J_\theta \end{array} \right)\left(\begin{array}{c} \delta\chi \\ \delta\theta \end{array} \right) + \left(\begin{array}{c} \delta\chi \\ \delta\theta \end{array} \right)^{\mathrm{T}} \left(\begin{array}{cc} G^{-1}_{\chi\chi} &\ G^{-1}_{\chi\theta} \\ G^{-1}_{\theta\chi} &\ G^{-1}_{\theta\theta} \end{array} \right) \left(\begin{array}{c} \delta\chi \\ \delta\theta \end{array} \right) \right]
\end{equation}We remind that $J_\chi$ and $J_\theta$ are not vanishing, since the point vortex ansatz is not a true solution of the GP equation.   After integrating over $\delta \chi$ and $\delta \theta$, we write $Z = \int \mathrm{d}\bm X_n \exp[\ii S_{\mathrm{eff}}[\bm X_n]]$ with \begin{equation}
S_{\mathrm{eff}} = S_{\mathrm{PV}}[\bm X_n] - \frac{1}{2}\int \mathrm{d}^3x\;\mathrm{d}^3x^\prime\; J_R(x) G_{RS}(x,x^\prime) J_S(x^\prime) + \frac{\ii}{2}\mathrm{tr}\;\log G
\end{equation}
where the $R,S,T$ indices denote the fields $\chi$ and $\theta$.

The classical contributions to the effective action, at one loop order, are equivalent to solving the Gross-Pitaevskii equation perturbatively, correcting the point vortex approximation at first order.   We will, for the remainder of this paper, ignore the $\mathrm{tr}\log$ term, as we are not interested in corrections to the effective action arising from quantum fluctuations.   In Appendix \ref{quantumapp} we sketch out how this quantum determinant can be computed, and point out that \begin{equation}
\frac{S_{\mathrm{classical}}}{S_{\mathrm{quantum}}} \sim \rho_0\xi^2 \equiv \mathcal{N} \label{quantumeq}
\end{equation}where $\mathcal{N}$ denotes the number of vortices absent from the background condensate at the superfluid core.   Thus, we see that quantum corrections are $1/\mathcal{N}$ (number fluctuation) suppressed, and can be systematically neglected in a limit where a superfluid vortex, made up of many atoms, is well defined.   In cold atomic gases, the s-wave scattering length is usually rather small, implying that $\lambda$ is small and thus that $\mathcal{N}\gg 1$ is a reasonable limit to consider \cite{fetter}.

\subsection{Green's Functions}
As we already know exactly how $J_{\chi,\theta}$ depend on $\bm X_n$, we simply have to compute the matrix $G$.  Note that, in the equations which follow, we do \emph{not} assume that the $\bm X_n(t)$ are on-shell, but we do assume that $\chi$ and $\theta$ take on the form of Eq. (\ref{eq13}).  We can straightforwardly compute (note that all derivatives act to the right)
\begin{subequations}\begin{align}
G^{-1}_{\theta\theta} &= \frac{\delta^2 S}{\delta \theta \delta \theta } = \nabla \cdot \left(\frac{\mu \ee^{2\chi}}{m\lambda}\nabla\right)\\
G^{-1}_{\theta\chi} &= \frac{\delta^2 S}{\delta \theta \delta \chi} = \left(\partial_t + \bm v \cdot \nabla\right) \frac{2\mu \ee^{2\chi}}{\lambda} \\
G^{-1}_{\chi\theta} &= \frac{\delta^2 S}{\delta \chi \delta \theta} =  -\frac{2\mu \ee^{2\chi}}{\lambda} \left(\partial_t + \bm v \cdot \nabla\right)   \\
G^{-1}_{\chi\chi} &= \frac{\delta^2 S}{\delta \chi \delta \chi} = -\frac{4\mu \ee^{2\chi}}{\lambda}\left(\partial_t\theta + \frac{(\nabla \theta)^2}{2m} + \mu \left(2\ee^{2\chi}-1\right)\right) + \frac{2\mu \ee^{2\chi}}{\lambda m} \nabla^2\chi + \frac{2\mu \ee^{2\chi}}{\lambda m}\nabla\chi\cdot\nabla + \frac{\mu \ee^{2\chi}}{\lambda m} \nabla^2
\end{align}
\end{subequations}
We can obtain $G$ by solving the equations of motion \begin{equation}
G^{-1}_{RS} G_{ST}(x,x^\prime) = -\delta_{RT} \delta(x-x^\prime).
\end{equation}

In the dilute limit, away from vortex cores, the contributions to $G$ from vortices are suppressed by powers of $\rstar$;  the leading order contributions to the Green's functions are simply those that arise in vacuum.  Thus, let us compute $G$ in the vacuum with no vortices.  In this case, we can Fourier transform $G^{-1}$ to obtain
\begin{equation}
G^{-1} = \left(\begin{array}{cc} -\mu k^2/m\lambda &\ -2\ii\omega \mu/\lambda \\ 2\ii \omega\mu/\lambda &\ -4\mu^2/\lambda-\mu k^2/\lambda m  \end{array}\right),
\end{equation}
which we can straightforwardly invert:
\begin{equation}
G = \frac{1}{4\mu^3m^{-1}\lambda^{-2}k^2(1+\xi^2k^2/2) -4(\omega \mu/\lambda)^2} \left(\begin{array}{cc} -4\mu^2/\lambda-\mu k^2/\lambda m &\ 2\ii \omega\mu/\lambda \\  -2\ii\omega \mu/\lambda &\    -\mu k^2/m\lambda \end{array}\right).
\end{equation}
In the long wavelength limit, this Green's function describes the propagation of simple sound waves, although it is not so transparent in this language.   By either computing directly from the equations of motion the dispersion relation $\omega(k)$ for the propagating waves, or by simply noticing that $G$ has a pole whenever
\begin{equation}
\omega^2 = c^2 k^2 \left(1+\frac{\xi^2k^2}{2}\right),  \label{dispeq}
\end{equation}
where the speed of sound is
\begin{equation}
c^2 = \frac{\mu}{m}
\end{equation}we obtain the dispersion relation Eq. (\ref{dispeq}).

The typical length scale involved in any (off-shell) fluctuation-mediated interaction between two vortices is of order $\rstar \gg \xi$.   In this limit, the on-shell dispersion relation is simply $\omega=\pm ck$.   The typical frequency scale involved is simply the typical velocity scale ($1/m\rstar$) divided by the length scale:  $1/m\rstar^2$.    It is now simple to check that $\omega_{\mathrm{sound}} \sim (\xi/\rstar) \omega_{\mathrm{on-shell}}$.    We conclude that if $\rstar\gg\xi$ it is appropriate to set $\omega \approx 0$ -- at leading order, vortices interact with each other instantaneously through ``virtual sound waves".   This is also the case in classical fluids \cite{nicolis}.

Setting $\omega=0$, the vacuum Green's function dramatically simplifies -- in fact, we can now exactly compute all of its components (in the relevant limit): \begin{subequations} \label{vacgreen} \begin{align}
G_{\theta\theta}(\bm x, \bm x^\prime) &\approx -\frac{\lambda m}{2\pi \mu} \log \frac{|\bm x - \bm x^\prime|}{\xi} ,  \\
G_{\chi\chi}(\bm x, \bm x^\prime) &\approx -\frac{\lambda}{4\mu^2}\delta(\bm x - \bm x^\prime).
\end{align}\end{subequations}
Of course, this is no longer true if either $\bm x$ or $\bm x^\prime$ is within a distance of order $\xi$ from any of the vortex cores.  In this case translation and rotation invariance will be strongly broken.   However, the vortex core itself simply proves a \emph{UV cut-off} to the effective theory.   Therefore, we simply model the presence of cores by truncating the Green's functions at distances of order $\xi$ from any vortex core.

\subsection{Outline and Summary of Results}
Let us now outline the computation of the effective action.  We need only keep the terms in $J_R$ which are the largest order in $\xi/\rstar$.  Based on our calculations in Appendix \ref{checkapp}, these turn out to be \begin{subequations}\begin{align}
J_\theta &\approx \frac{2\mu}{\lambda}\left[\partial_t \chi + \bm v \cdot \nabla \chi\right], \\
J_\chi &\approx -\frac{2\mu}{\lambda}\left[\partial_t\theta + \frac{(\nabla\theta)^2}{2m}\right].
\end{align}\end{subequations}

We have relegated details of the resulting integrals to Appendix \ref{dimreg}.   Although some integrals in the effective action cannot be done exactly, we are able to extract leading order asymptotic behaviors when appropriate.

We find that \begin{equation}
\mathcal{S}_\theta \equiv -\frac{1}{2}\int \mathrm{d}^2\bm x \mathrm{d}^2\bm x^\prime G_{\theta\theta}(\bm x, \bm x^\prime) J_\theta(\bm x)J_\theta(\bm x^\prime) \sim \frac{\rho_0}{\mu} \sum_n \left(\bm U_n - \dot{\bm X}_n\right)^2.   \label{eqeff1}
\end{equation}This contribution to the equations of motion is a sum of squares, each which vanishes on the point-vortex ansatz.   This means that when we vary $\mathcal{S}_\theta$ to compute the equations of motion for $\bm X_n$, these contributions will not contribute -- it is consistent at this order (in the equations of motion) to set $\dot{\bm X}_n = \bm U_n$.

The $J_\chi$ contributions are a bit more lengthy, and are outlined in the appendix.    The integrals we have to evaluate are \begin{equation}
\mathcal{S}_\chi \equiv -\frac{1}{2}\int \mathrm{d}^2\bm x \mathrm{d}^2\bm x^\prime G_{\theta\theta}(\bm x, \bm x^\prime) J_\theta(\bm x)J_\theta(\bm x^\prime) = \frac{m^2}{2\lambda}\int \mathrm{d}^2\bm x \left(\sum_n\bm v_n \cdot \dot{\bm X}_n - \sum_{m\ne n} \bm v_m \cdot \bm v_n \right)^2. \label{eqeff2}
\end{equation}There are a variety of contributions involving products of velocities over 2, 3 and 4 distinct vortices, respectively.   We provide more details in the appendix.

Many of the integrals over products of velocities, obtained by expanding out Eq. (\ref{eqeff2}), can have logarithmic divergences in the IR and/or the UV.  Remarkably, when we sum together all of these logarithmic divergences, these divergent contributions to the action greatly simplify: \begin{equation}
S = S_{\mathrm{PV}} + \int \mathrm{d}t \; \frac{\rho_0\pi}{2\mu}\left(\log\frac{L}{r_{\mathrm{typ}}} \left(\sum_m \Gamma_m \dot{\bm{X}}_m\right)^2 + \log \frac{r_{\mathrm{typ}}}{\xi} \sum_{m}\left(\dot{\bm{X}}_m-\bm U_m\right)^2\right) + \cdots \label{eq41}
\end{equation}In this equation, we are using $r_{\mathrm{typ}}$ rather carelessly to denote a cutoff-independent distance between two vortices. Our main point is as follows.   Suppose that we keep an identical configuration of point vortices, but shrink $\xi$ by a factor of $\lambda$.   The correction to the action is proportional to a sum of squares, where \emph{each object being squared vanishes on point-vortex dynamics}.   Therefore, when we take a variational derivative and obtain the equations of motion, the equations of motion are independent of the redefinition of $\xi$ by the factor $\lambda$.   An identical argument holds for $L$ -- we note that $\sum \Gamma_m \dot{\bm{X}}_m = \bm 0$ on point vortex dynamics, following Eq. (\ref{momcons}).   Because of these cancellations, we see that \emph{no logarithmic divergences alter the equations of motion}.

This is in contrast to the most common argument (see, e.g., \cite{thompson}), which states that the vortex mass in the equations of motion is logarithmically divergent as $\log(L/\xi)$.   Indeed, we find this result in Eq. (\ref{eq74}).   However, multiple logarithmic divergences conspire to exactly cancel in the equations of motion, at leading order.   Of course, these logarithmic divergences may in principle be important at higher orders away from point vortex dynamics, but calculations based on perturbation theory cannot be trusted in this regime without a systematic consideration of all possible second-order perturbations.   In fact, even for a single vortex, we should exercise caution assuming that vortex mass is logarithmically divergent -- $\dot{\bm X} = \bm 0$ for a single vortex, and thus one needs to perform a higher order perturbative calculation in an external superfluid velocity field in order to determine the corrections to the vortex trajectory.


\section{Dynamics of Two Vortices}  \label{smokesection}
We have already seen that the only contribution to the action at $\mathcal{O}(\rstar^{-2})$ comes from the $J_\chi^2$ term: \begin{equation}
\mathcal{S}_\chi = \frac{m^2}{2\lambda} \int \mathrm{d}^3x \left( \bm v_1 \cdot \bm v_2 - \bm v_1 \cdot{\dot{\bm X}}_1 - \bm v_2 \cdot \dot{\bm{X}}_2\right)^2.
\end{equation}
Using dimensional regularization with minimal subtraction \cite{peskin} we are able to compute these integrals exactly, as we carefully show in Appendix \ref{dimreg}: \begin{align}
\mathcal{S}_\chi &= \frac{\pi \rho_0}{2\mu} \int \mathrm{d}t  \left[ \log\frac{L}{|\bm X_1-\bm X_2|}\left(\Gamma_1 \dot{\bm{X}}_1 + \Gamma_2 \dot{\bm{X}}_2\right)^2   \right.
+ \log \frac{|\bm X_1 - \bm X_2|}{\xi}\left((\dot{\bm{X}}_1 - \bm U_1)^2 +( \dot{\bm{X}}_2-\bm U_2)^2  \right)  \notag \\
& \left. -2 \Gamma_1\Gamma_2\dot{X}_1^i\dot{X}_2^j\epsilon^{ik}\epsilon^{jl}\frac{(X_1-X_2)^k(X_1-X_2)^l}{|\bm X_1-\bm X_2|^2} + \frac{1}{m^2|\bm X_1-\bm X_2|^2}  \right]  \label{smokering}
\end{align}The equations of motion from this action can be written in a rather simple form, when we use that $\dot{\bm{X}}_1 \approx \bm U_1$ and $\dot{\bm{X}}_2 \approx \bm U_2$ to simplify the corrections to the equations of motion once we are on-shell:
\begin{align}
\epsilon_{ij}\Gamma_1 \left(\dot{X}_1^j - U_1^j\right) + \frac{1}{\mu}\ddot{X}_1^i + \frac{2\xi^2 (X_1-X_2)^i}{m|\bm X_1-\bm X_2|^4} &= \notag \\
 \epsilon_{ij}\Gamma_1 \left(\dot{X}_1^j - U_1^j\right) + \frac{1}{\mu}\ddot{X}_1^i - \frac{m^2}{\mu}\Gamma_2 \dot{\bm{X}}_1\cdot\dot{\bm{X}}_1 \epsilon_{ij} \dot{X}_1^j  &= 0.  \label{eff2}
\end{align}and an identical equation if we swap 1 and 2.   We thus find the equations of motion for this pair take on a deceptively simple form.   (We have no reason to expect so many cancellations to occur for the more general case of $N$ vortices.)

\subsection{Vortex-Antivortex Pair}

We can now use this effective equation of motion to argue that a pair of vortices with $\Gamma_1 = -\Gamma_2=1$ cannot annihilate as long as their distance apart $r_0 \gg \xi$, and the dynamics is only perturbed weakly from point vortex dynamics.     The reason for this is very simple.   There are conserved quantities associated with Eq. (\ref{eff2}) which we can interpret as energy and momentum: \begin{subequations}\begin{align}
E &= \pi\rho_0 \left[\frac{1}{m}\log \frac{|\bm X_{1}-\bm X_2|}{\xi} - \frac{\xi^2}{m|\bm X_{12}|^2} + \frac{\dot{\bm{X}}_1^2 + \dot{\bm{X}}_2^2}{2\mu}\right], \\
P^i &= \pi\rho_0 \left[\epsilon^{ij}(X_1-X_2)^j + \frac{\dot{X}^i_1 + \dot{X}^i_2}{\mu} \right].
\end{align}\end{subequations}Writing $\Delta \bm X = \bm X_1 - \bm X_2$, it is simple to see that there is a conserved quantity \begin{equation}
\frac{E}{\pi\rho_0} = \frac{1}{m}\log \frac{|\Delta \bm X|}{\xi} - \frac{\xi^2}{m|\Delta \bm X|^2} + \frac{\Delta\dot{\bm{X}}^2}{4\mu}  +\frac{\mu}{4} \left(\epsilon^{ij}\frac{P^j}{\pi\rho_0} + \Delta  X^i\right)^2.
\end{equation}
For simplicity, let us assume that at time $t=0$ we have the $\Gamma_1=+1$ vortex at $x=r_0/2$, and the $\Gamma_2=-1$ vortex at $x=-r_0/2$.   Both vortices have $y=0$ at this time, and have $\dot{Y}_{1,2} = -1/mr_0$ on the point-vortex ansatz; let us assume that at $t=0$, the velocity of both vortices is given by point-vortex dynamics.   Let us denote $\Delta X = r_0 + \delta x$, and $\Delta Y = \delta y$.   Furthermore, using the value of $\bm P$ on the point vortex ansatz -- including the subleading term due to vortex velocity in $\bm P$ -- we obtain
\begin{equation}
\frac{E}{\pi\rho_0} \approx \text{constant} +  \frac{2\xi^2}{mr_0^3}\delta x + \frac{\mu}{4}\left(\delta x^2 + \delta y^2\right) + \frac{\delta\dot{x}^2+\delta\dot{y}^2}{4\mu} + \mathcal{O}(\delta r^3)
\end{equation}The dynamics of $\delta x$ and $\delta y$ are well-described by simple harmonic oscillation about $\delta x \approx -4\xi^4/r_0^3$.   Thus, weak perturbations will not grow -- the vortex-antivortex pair will always be a distance $\approx r_0$ apart, and will never annihilate.   Importantly, note that the sound-induced velocities are of order $\delta \dot{r} \sim \mu \delta r \sim (1/mr_0 )(\xi^2/r_0^2)$, ensuring that we do not exit the regime of validity of our effective theory.    Our conclusions are unchanged if the initial velocities do not exactly coincide with point vortex dynamics, but only differs by a factor $\sim 1/r_0^3$ -- consistent with the order at which the equations of motion of point vortex dynamics is corrected.

\section{Turbulence}
Let us now extract some lessons from our computation for two-dimensional superfluid turbulence at zero temperature.    The discussion below is non-rigorous, and it would be worthwhile to check these claims carefully (most likely by simulation) in the future.

To begin with, our effective action for vortices alone has an exactly conserved energy $E$, which follows directly from the time translation invariance of the action.    This conserved energy does not have any obvious pathologies: for example, in the case where we can do the exact calculation in the previous section, the ``mass term" in $E$ has a positive coefficient.   Due to radiation of sound waves by accelerating vortices \cite{barenghi1, barenghi2}, the energy stored in vortices alone may not stay conserved at the next order in the calculation.   This follows from the fact that the power radiated  scales as $\sim \ddot{X}^2\sim \rstar^{-6}$, comparable to the rate of change $\sim \rstar^{-2}$ of the corrections to the energy at second order in sound-induced corrections, which will be $\Delta E \sim \rstar^{-4}$.   This is a second order correction ($\sim \xi^4/\rstar^4$ relative to the leading order point vortex dynamics) and is quite subleading in a dilute limit, so we do not consider it further -- the energy stored in vortices is conserved to very good approximation.

Next, we can consider a thought experiment where the sound-induced corrections to vortex motion have brought a vortex-antivortex pair within a distance $r_2 \ll \rstar$; these two vortices are far closer to each other than to any other vortex in the system.  If $r_2 \gg \xi$ as well, then we can use our effective theory to analyze the dynamics.    Analyzing the behavior of the 3 and 4 velocity integrals in $\mathcal{S}_\chi$, we find that the terms in the action that couple this pair of vortices to the remainder of the system are suppressed by a factor of $r_2/\rstar$.    At leading order in $r_2/\rstar$, the dynamics of the previous section are applicable;   the combination of a conserved vortex energy and the decoupling of the dynamics of vortex pairs as they approach each other is suggestive that, at leading order in the dilute limit, there is no sound-induced instability to vortex annihilation with $N>2$ vortices.

Of course, our effective theory breaks down at $\rstar \sim \xi$, and  in this limit vortex annihilation and creation events have been observed numerically at zero temperature \cite{simula, numasoto}.   We also note that the vortex annihilation observed in the recent experiment \cite{becexp} is at finite temperature; \cite{lucas} describes the effective theory relevant for vortex annihilation in this experiment, which allows for vortex annihilation events even at leading order.

If instabilities to vortex annihilation are not easily created by sound-mediated vortex interactions, then we conclude that point vortex dynamics is a legitimate description of superfluid dynamics at zero temperature, at leading order in $\xi/\rstar$.  The velocity fields of point vortex dynamics exactly satisfy the Euler equation of a normal fluid.  In this sense, the dynamics of large clusters of point vortices is rather similar to the dynamics of a continuous (normal) two-dimensional fluid \cite{onsager, siggia}.    Turbulent flows in a two-dimensional normal fluid are known to be characterized by an inverse cascade of energy to long wavelengths \cite{kraichnan}.   Even though the precise nature of forcing or of dissipation in the superfluid -- at zero temperature, the emission of sound waves --  is distinct from viscous dissipation in the normal fluid, over the great majority of length scales, the superfluid is described by an identical equation to a classical normal fluid.   Over the inertial range of classical turbulence, the Euler equation is the effective equation governing the dynamics.   It is reasonable that a dilute zero temperature vortex liquid in a superfluid behaves similarly to a classical fluid and undergoes an inverse energy cascade, by which vortices organize themselves into large scale structures.

\section{Conclusion}
In summary, we have constructed the next-to-leading-order effective action describing quantized vortices of winding number $\pm 1$ in a two-dimensional superfluid described by GP theory.   We found that the corrections extend far beyond a simple mass term -- indeed, a plethora of new terms, consistent with the symmetries of the problem and the long-range tails of vortex velocities, appear at next-to-leading order.    Indeed, the rather complicated nature of the answer is suggestive that much of it is non-universal to different actions, though we would not be surprised if the cancellation of all logarithmic divergences (on-shell) at next-to-leading order was a generic feature of reasonable theories.

We were able to exactly determine the next-to-leading-order action for a \emph{pair} of vortices.  A careful analysis of the vortex-antivortex pair demonstrated no instability to annihilation.    We argued that this lack of instability implies that turbulent superfluids of dilute vortices, at zero temperature, should be well-described by point vortex dynamics.    Point vortex dynamics itself provides a discretized approximation to the dynamics of a classical continuum fluid.   This is suggestive that the inverse cascade picture of classical turbulence is qualitatively correct for a turbulent zero temperature superfluid.

We stress that the discussion above about inverse cascades in superfluid turbulence are only valid at zero temperature.   There are zeroth-order corrections to the vortex equations of motion at finite temperature.   The consequences of finite temperature on superfluid turbulence are discussed in \cite{lucas}.

\addcontentsline{toc}{section}{Acknowledgements}
\section*{Acknowledgements}
We would like to thank Gregory Eyink, Gregory Falkovich, V.P. Nair, Achim Rosch, Subir Sachdev, Philipp Strack, Tadashi Tokieda, Marija Vucelja and Will Witczak-Krempa for helpful discussions. We especially thank Paul Chesler for many useful discussions and collaboration for much of this work.

AL is supported by the Smith Family Graduate Science and Engineering Fellowship.  AL would like to thank the Perimeter Institute of Theoretical Physics for hospitality during latter stages of this work.  Research at Perimeter Institute is supported by the Government of Canada through Industry Canada and by the Province of Ontario through the Ministry of Economic Development \& Innovation. PS would like to thank the Simons Center for Geometry and Physics during the program ``Quantum Anomalies, Topology, and Hydrodynamics'' for hospitality and partial support. PS was supported by a Marie Curie International Outgoing Fellowship, grant number PIOF-GA-2011-300528.

\begin{appendix}
\titleformat{\section}
  {\gdef\sectionlabel{}
   \Large\bfseries\scshape}
  {\gdef\sectionlabel{\thesection. }}{0pt}
  {\begin{tikzpicture}[remember picture,overlay]
	\draw (-0.2, 0) node[right] {\textsf{Appendix \sectionlabel#1}};
	\draw[thick] (0, -0.4) -- (\textwidth, -0.4);
       \end{tikzpicture}
  }
\titlespacing*{\section}{0pt}{15pt}{20pt}

\section{Point Vortex Dynamics in the Dilute Limit}\label{checkapp}
In this appendix, we carefully analyze the equations of motion, and show that everywhere in space, point vortex dynamics is a good approximation.   First we analyze the $J_\theta$ term, which encodes conservation of particles.   Using that $\nabla \cdot \bm v_n = 0$, we can simplify Eq. (\ref{eqjtheta}) to
\begin{equation}
J_\theta = \frac{2\mu \ee^{2\chi}}{\lambda} \sum_n \nabla \chi_n \cdot \left(- \bm { \dot{X}}_n + \bm v \right)   \label{jtheta}
\end{equation}  Note that $J_\theta$ can be written as the divergence of a vector.
We now need to analyze the sizes of various terms in this equation.  For a distance $r\lesssim \xi$ to vortex core $n$, the largest terms in $J_\theta$ are $\ee^{2\chi} \bm {\dot{X}}_n \cdot \nabla \chi_n $ and $\ee^{2\chi} \bm v \cdot \nabla \chi_n $.    Exactly at the vortex core, these two terms cancel, since $\bm v_n \cdot \nabla \chi_n = 0$ and $\dot{\bm X}_n$ is given by the local superfluid velocity through the core.   A distance $\sim \xi $ away from the vortex, one finds that, as $\nabla \chi_n \sim 1/\xi$, and $ (\bm {\dot{X}}_n - \bm v)(|\bm x - \bm X_n|)  \sim \xi / m\rstar^2$: \begin{equation}
J_\theta \sim (\bm {\dot{X}}_n - \bm v) \cdot \nabla \chi_n  \sim \frac{1}{m\rstar^2}.  \label{eq49}
\end{equation}A distance $\sim \xi$ away from the vortex core, the leading order contributions to $J_\theta$ are $\bm v\cdot \nabla \chi_n$, which is singular;  Eq. (\ref{eq49}) is asymptotically small compared to these terms, and so point vortex dynamics is a good approximation.    Away from vortex cores, we find that $ \dot{\bm X}_n$ is uncorrelated with $\bm v$, but $\nabla \ccore(\bm x- \bm X_n) \sim \xi^2/\rstar^3$, so we conclude that
\begin{equation}
J_\theta \sim \sum_n (\bm {\dot{X}}_n - \bm v)\cdot \nabla \chi_n \sim \sum_n \frac{1}{\rstar} \frac{1}{r_n^3} \sim \frac{1}{\rstar^4}
\end{equation}
Note that we have used an averaging argument in the last step in the above equation, analogous to Eq. (\ref{averageeq}) -- we will employ this frequently.   Everywhere in space, we see that $J_\theta$ is suppressed by a power of $\rstar$, which means that point vortex dynamics is exact in the $\rstar \rightarrow \infty$ limit.   In particular, the largest correction to $J_\theta$ occurs for $r\lesssim \xi$.

Next, let us analyze the $J_\chi$ term.    It is helpful to first subtract out all of the terms which are vanishing due to the fact that $\chi_0$ exactly solves the Gross-Pitaevskii equation for a single vortex:
\begin{align}
J_\chi = - \frac{2\mu \ee^{2\chi}}{\lambda} &\left\lbrace  \sum_n\left[\frac{(\nabla\theta_n)^2}{2m} + \mu \left(\ee^{2\chi_n}-1\right) - \frac{\nabla^2\chi_n + (\nabla\chi_n)^2}{2m}\right] \right. \notag \\
&+  \left. \sum_n \partial_t\theta_n + \mu \left(\ee^{2\chi}-1\right) - \sum_n \mu \left(\ee^{2\chi_n}-1\right) + \sum_{m\ne n} \frac{\nabla \theta_m\cdot\nabla\theta_n - \nabla \chi_m\cdot\nabla\chi_n}{2m}    \right\rbrace
\end{align}
We have written $J_\chi$ in this expanded form for a reason:  the terms in the first line exactly cancel each other, because $\chi_0$ is an exact solution to the Gross-Pitaevskii equation.   It remains to analyze the terms on the second line.   To analyze the size of the remaining terms, we first analyze the non-vanishing terms coming from the pressure $P$:
\begin{equation}
m\Delta P = \frac{1}{2m}\sum_{k\ne l} \nabla \chi_k \cdot \nabla \chi_l + \mu \left(\ee^{2\chi}-1\right) - \mu \sum_l \left(\ee^{2\chi_l}-1\right).
\end{equation}
Near vortex core $k$, it turns out that the most divergent term in the second line is $\nabla \chi_k \cdot \nabla \chi_l /2m \sim \xi^2/(mr\rstar^3)$.   Although this is appears divergent, we must remember that  every term is multiplied by $\ee^{2\chi} \sim (r/\xi)^2$, and so in fact this term is vanishing $\sim r$.   The most divergent terms, which cancel on point vortex dynamics, come from the first line, and scale as $\ee^{2\chi}/r^2 \sim r^0$: evidently, point-vortex dynamics is still a good approximation.   The remaining pressure terms are rather small, as we can separate out the sum over $l$ into a sum over $l\ne k$ (these terms scale as $\xi^2/\rstar^2$) and then combine the remaining $\mu$-terms:  $\mu(\ee^{2\chi}-\ee^{2\chi_k}) = \mu \ee^{2\chi_k} (\exp[\sum_{l\ne k} \chi_l] - 1) \sim \mu (r/\xi)^2$ near a vortex core.   Thus the first term dominates near a vortex core.    Away from the vortex cores, we have $\nabla \chi_k \cdot \nabla \chi_j \sim \xi^4/\rstar^6$.   The non-derivative terms are approximately given by
\begin{equation}
\mu \left(\ee^{2\chi}-1\right) - \sum_l \mu \left(\ee^{2\chi_l}-1\right) \approx 2\mu\sum_{l\ne k} \chi_l \chi_k,
\end{equation}
which scales as $\mu (\xi/\rstar)^4$.

There are also non-vanishing contributions to $J_\chi$ coming from the time derivative and velocity terms: $\partial_t \theta + (m/2) \sum_{l\ne k} \bm v_l \cdot \bm v_k$.    Near vortex core $k$, we rearrange this sum as \begin{equation}
\partial_t \theta + \frac{m}{2}\sum_{l\ne k} \bm v_l \cdot \bm v_k = \sum_k \left(\sum_{l\ne k} \bm v_l - \dot{\bm X}_k\right)\cdot \bm v_k - m\sum_{l\ne k} \bm v_l \cdot \bm \dot{X}_l + \frac{m}{2}\sum_{l\ne n\ne k}\bm v_n \cdot \bm v_l.   \label{eq54}
\end{equation} The last two terms $\sim 1/\rstar^2$.   The object in parentheses in the first term vanishes at vortex cores, and $\sim \Delta x/\rstar^2$ away from vortex cores, and since $\bm v_k \sim 1/\Delta x$ near the vortex core, we see that there is no singularity and all terms scale as $1/\rstar^2$.   The leading order terms (e.g. $\partial_t \theta_n$) are singular, and so all corrections are subleading.  Eq. (\ref{eq54}) are the largest corrections, dominating over $\chi$-induced corrections.

The overall factor of $\ee^{2\chi}$ in front of $J_\chi$ induces corrections that are more subleading.  We conclude point vortex dynamics is an approximate solution to the equation of motion, up to subleading corrections of order $\xi^2/\rstar^2$.

\section{Galilean Invariance}\label{galilean}
\subsection{Gross-Pitaevskii Equation}
Here we review, for convenience, how the Gross-Pitaevskii action (and equations of motion) are Galilean invariant.   This symmetry follows most naturally after writing out $\psi = \sqrt{\rho_0} \ee^{\chi + i\theta}$, as we have in the main text.  As $\nabla \theta$ corresponds to the superfluid velocity, and we expect that $\nabla \theta \rightarrow \nabla \theta + \bm v$ under a Galilean boost of velocity $\bm v$, one can easily check that $\chi$ is invariant under a Galilean boost, and
\begin{equation}
\theta \rightarrow \theta + \bm v \cdot \bm x - \frac{\bm v^2}{2m} t
\end{equation}
leaves the action Eq. (\ref{eq5}) invariant.
\subsection{Point-Vortex Dynamics}
The Lagrangian of point vortex dynamics as written down in the main text is not Galilean invariant.   There is a physical reason for this.   The presence of a single vortex ``breaks Galilean invariance" by picking out a preferred rest frame -- namely, the one where the vortex is at rest.  Of course, there is an equivalent description of the physics in a frame moving at a relative velocity $\bm V$ -- in this case, both the vortex and the superfluid at spatial infinity are moving at a constant velocity $\bm V$.

An analogous story holds for the point vortex action in the main text.   The preferred rest frame we have chosen corresponds to a frame in which the superfluid velocity at spatial infinity is zero.   To make the equations of motion of point-vortex dynamics Galilean invariant, we simply must modify the equations of motion to \begin{equation}
\dot{\bm{X}}_m - \bm V = \bm U_m(\bm X_n)
\end{equation}Here $\bm V$ is an auxiliary non-dynamical variable, corresponding to the velocity of the superfluid at infinity.   Galilean invariance has been restored if we transform $\bm X_m \rightarrow \bm X_m + \bm a t$, $\bm V \rightarrow \bm V + \bm a$.   This parameter $\bm V$ appears in the action as \begin{equation}
S = \int \mathrm{d}t \left[ \sum_n \frac{\Gamma_n}{2}\epsilon_{ij}\left(\dot{X}^i -2V^i\right)X^j + \sum_{m\ne n} \frac{\Gamma_m\Gamma_n}{m} \log \frac{|\bm X_m - \bm X_n|}{L}\right].
\end{equation}

\section{Quantum Corrections}\label{quantumapp}
In this short appendix we point out why quantum corrections to the effective action can be neglected.

Let us begin by computing the typical scale of $J_R G_{RS} J_S$.   As we argue in the main text, the dominant contributions to the classical corrections to the equations of motion come from the $J_\chi$ integral: \begin{equation}
S_{\mathrm{classical}} \sim \int \mathrm{d}^2\bm x \mathrm{d}t \; \frac{\lambda J_\chi^2 }{\mu^2} \sim \int \mathrm{d}^2\bm x \mathrm{d}t\;  \frac{\rho_0}{\mu (m\rstar^2)^2} .
\end{equation}In our scaling argument, we have focused on the scaling of $J_\chi$ in the middle of the vortex cloud, where most of the contributions to the effective action arise.

Now let us look at the quantum corrections.   At leading order, we can approximate that \begin{equation}
\mathrm{tr} \; \log G^{-1} = \mathrm{tr} \; \log \left(G_0^{-1} \left(1+G_0 \delta G^{-1}\right)\right) \approx \mathrm{tr}\; \log G_0^{-1} + \mathrm{tr}\; \left(G_0 \delta G^{-1} - \frac{(G_0\delta G^{-1})^2}{2}\right)
\end{equation}
where $G_0$ is the vacuum Green's function, and $\delta G^{-1}$ is the corrections to the inverse Green's function.   We must go to second order in $\chi$ to obtain an answer which will not vanish by translation invariance -- i.e., so that our trace contains the product $\chi_m\chi_n$, e.g.  These terms can be analyzed with similar scaling arguments.   For simplicity, let us focus on a single example of terms which arise in the $\delta G^{-1}$-dependent contribution: \begin{equation}
\mathrm{tr}\log \left(G_{0\chi\chi} \delta G^{-1}_{\chi\chi} - \frac{(G_{0\chi\chi} \delta G^{-1}_{\chi\chi})^2}{2}\right) \sim \int \mathrm{d}^2 \bm x \mathrm{d}^2\bm x^\prime  \mathrm{d}t  \mathrm{d}t^\prime \; \chi_m \chi_n \left(\delta(\bm x- \bm x^\prime)\delta(t-t^\prime)\right)^2.
\end{equation}Now, recall that our theory comes endowed with natural cutoffs:  the length scale $\xi$ for a UV cutoff on $\bm x$ integrals, and the time scale $m\xi^2$ for $t$ integrals.  This allows us to make sense of the square of a $\delta$ function by replacing, e.g., $\int \mathrm{d}t^\prime \delta(t-t^\prime)^2 \sim 1/m\xi^2$.  We conclude that \begin{equation}
S_{\mathrm{quantum}} \sim \int \mathrm{d}^2\bm x \mathrm{d}t \; \frac{\xi^4}{\rstar^4} \frac{1}{m\xi^4}
 \sim \int \mathrm{d}^2\bm x \mathrm{d}t \; \frac{1}{\mu  \xi^2 (m\rstar^2)^2}
\end{equation}
All other terms in $S_{\mathrm{quantum}}$ can be shown to scale similarly.   It is now straightforward to observe that the quantum contributions to the effective action are suppressed by a factor of $\mathcal{N}$, as we stated in Eq. (\ref{quantumeq}).

\section{Computation of the Effective Action}\label{dimreg}
In this appendix we discuss the evaluation of the integrals involved in the main text.

\subsection{The $\mathcal{S}_\theta$ Integrals}
Using Eq. (\ref{jtheta}), the fact that  $\bm v_n \cdot \nabla \chi_n =0$, $\nabla \cdot \bm v = 0$ and integrating by parts we find that
\begin{equation}
\mathcal{S}_\theta = \frac{2}{\lambda \xi^2} \int \mathrm{d}^2 \bm x \mathrm{d}^2\bm x^\prime \left[\partial_i \partial_j^\prime \frac{\log |\bm x- \bm x^\prime|}{2\pi}\right] \sum_{m,n} \chi_m(\bm x) \chi_n(\bm x^\prime) \left(\sum_{l\ne m} \bm v_l(\bm x) - \dot{\bm{X}}_m\right)_i \left(\sum_{p\ne n} \bm v_p(\bm x^\prime) - \dot{\bm{X}}_n\right)_j
\end{equation}We can now exploit the fact that if $m\ne n$, $\chi_m \chi_n$ is always suppressed by a factor of $\rstar^{-2}$ -- the dominant contribution from this sum necessarily comes from the sum over $m=n$.    Note that $\chi_m \chi_n \approx \xi^4/|\bm x|^2 |\bm x^\prime|^2$, and \begin{equation}
\partial_i \partial_j^\prime \frac{\log |\bm x- \bm x^\prime|}{2\pi} = \frac{\delta_{ij}}{2} \delta(\bm x - \bm x^\prime) + \frac{1}{2\pi|\bm x-\bm x^\prime|^2}\left(2\frac{(\bm x-\bm x^\prime)_i (\bm x-\bm x^\prime_j)}{|\bm x-\bm x^\prime|^2} - \delta_{ij}\right).
\end{equation}In particular, the key observation is that this integral is very sensitive to UV physics, but \emph{not} to IR physics, where the integrals converge, since the integrand falls off as $r^{-6}$.   In the UV, near vortex core $n$, we may approximate $\bm v_l $ ($l\ne n$) by a constant,  and in this case the only contribution to the integral comes from the $\delta$ function;  the other term vanishes by symmetry after integration with $\bm x$, $\bm x^\prime$.   To leading order in $(\xi/\rstar)^2$, \begin{align}
\mathcal{S}_\theta  \approx \frac{1}{4\lambda} \sum_{n} \int \mathrm{d}^2\bm x \; \frac{\xi^2}{|\bm x-\bm X_n|^4} \left(\bm U_n - \dot{\bm{X}}_n \right)^2 \sim \sum_n \frac{\pi}{2\lambda} \left(\bm U_n - \dot{\bm{X}}_n\right)^2.
\end{align}The last step in this integral is extremely sensitive to the nature of the UV cut-off -- the multiplicative constant sitting in front of the integral is sensitive to near-core physics.   Thus, we have indicated our ignorance of the overall coefficient of this term with the $\sim$ symbol.

\subsection{The $\mathcal{S}_\chi$ Integrals}
Next, let us discuss the $J_\chi^2$ contributions to the effective action.     We only need to consider, as with $J_\theta^2$, the terms which do not vanish on the vortex ansatz.  In particular, the contributions to $J_\chi^2$ due to fluctuations in $\chi$ will necessarily be suppressed by a factor of at least $\xi^2/\rstar^2$.\footnote{Recall that single $\chi_n$ terms in $J_\chi$ exactly cancel -- the only terms involving $\chi$ which do not vanish on point vortex dynamics involve either $\chi_m \chi_n$, $\chi_m (\nabla \theta)^2$, or $\chi_m \partial_t\theta$, each of which is $\sim \rstar^{-4}$.}   The only terms which contribute at this order are \begin{equation}
-\frac{1}{2}\int \mathrm{d}^2\bm x \mathrm{d}^2\bm x^\prime \; G_{\chi\chi}(\bm x, \bm x^\prime) J_\chi(\bm x)J_\chi(\bm x^\prime) \approx \frac{1}{2\lambda} \int \mathrm{d}^2\bm x \; m^2\left(\sum_{m\ne n} \frac{1}{2}\bm v_m \cdot \bm v_n - \sum_n \bm v_n \cdot \dot{\bm{X}}_n\right)^2.
\end{equation}

\subsubsection{2-Velocity Integrals}

First we describe thoroughly how to evaluate $\int \mathrm{d}^2\bm x\; v_1^i v_2^j$, which can be obtained via: \begin{equation}
I_{2,1ij} \equiv \int \mathrm{d}^2x \frac{(x-X_1)_i (x-X_2)_j}{(x-X_1)^2 (x-X_2)^2}.
\end{equation}We write \begin{equation}
I_{2,1ij} = \int \mathrm{d}^2x \int\limits_0^\infty \mathrm{d}s_1\mathrm{d}s_2 (x-X_1)_i (x-X_2)_j \exp\left[-s_1(\bm x-\bm X_1)^2 - s_2 (\bm x-\bm X_2)^2\right].
\end{equation}Defining $s_1 = S\alpha$, and $s_2=S(1-\alpha)$, as well as shifting the integral over $\bm x$ to \begin{equation}
\bm y = \bm x - \alpha \bm X_1  -(1-\alpha)\bm X_2,
\end{equation} we finally obtain \begin{equation}
I_{2,1ij} = \int \mathrm{d}^2y \int\limits_0^\infty \mathrm{d}S\int\limits_0^1 \mathrm{d}\alpha  \left(y -(1-\alpha)X_{12}\right)_i  (y+\alpha X_{12})_j \exp\left[-S\bm y^2 - S \alpha(1-\alpha)|\bm X_{12}|^2\right]
\end{equation}
with $\bm X_{12} \equiv \bm X_1 - \bm X_2$.  This integral is logarithmically divergent at long distances, and so we regulate it by continuing to $d=2-\epsilon$.   The factor $L_{\mathrm{DR}}$ here serves as an IR cut-off -- it may differ from the physical cutoff $L$ by a constant factor.   We then perform the Gaussian integrals to obtain \begin{align}
I_{2,1ij} &= L_{\mathrm{DR}}^{\epsilon} \int \mathrm{d}^{2-\epsilon}y \mathrm{d}S\mathrm{d}\alpha \; S  \left(y -(1-\alpha)X_{12}\right)_i  (y+\alpha X_{12})_j \exp\left[-S\bm y^2 - S \alpha(1-\alpha)|\bm X_{12}|^2\right] \notag \\
&= L_{\mathrm{DR}}^{\epsilon} \int \mathrm{d}S \mathrm{d}\alpha \; S \left(\frac{\pi}{S}\right)^{1-\epsilon/2} \left(\frac{\delta_{ij}}{2S} - \alpha(1-\alpha) X_{12i}X_{12j}\right) \exp\left[-S\alpha(1-\alpha)|\bm X_{12}|^2\right] \notag \\
&= \int\limits_0^1\mathrm{d}\alpha \left[ \frac{L_{\mathrm{DR}}^\epsilon \pi^{1-\epsilon/2}\delta_{ij}}{2 (\alpha(1-\alpha)|\bm X_{12}|^2)^{\epsilon/2}} \Gamma \left(\frac{\epsilon}{2}\right)- \pi \frac{X_{12i}X_{12j}}{|\bm X_{12}|^2} \right]  + \mathcal{O}(\epsilon).
\end{align}As the latter term is finite, we have already taken the $\epsilon\rightarrow 0$ limit.  The $\alpha$ integral in the first term can be explicitly evaluated:\begin{equation}
\left(\frac{L_{\mathrm{DR}}}{|\bm X_{12}|}\right)^\epsilon \Gamma\left(\frac{\epsilon}{2}\right)\frac{\pi^{1-\epsilon/2}}{2} \frac{\Gamma(1-\epsilon/2)^2}{\Gamma(2-\epsilon)} = \frac{\pi}{\epsilon} + \frac{\pi}{2}\left[2\log\frac{L_{\mathrm{DR}}}{|\bm X_{12}|} +2- \gamma - \log\pi \right]  + \mathcal{O}(\epsilon)
\end{equation}where $\gamma\approx 0.57$ is the Euler-Mascheroni constant.    We find \begin{equation}
I_{2,1ij} = \pi \delta_{ij} \left[ \frac{1}{\epsilon} +  \log \frac{L_{\mathrm{DR}}}{|\bm X_{12}|} +  1 - \frac{\gamma+\log\pi}{2} \right] - \pi \frac{X_{12i}X_{12j}}{|\bm X_{12}|^2}.  \label{i21}
\end{equation}

Next we evaluate $\int \mathrm{d}^2\bm x\; v_1^i v_1^j$: \begin{equation}
I_{2,0ij} \equiv \int \mathrm{d}^2\bm x \frac{x_ix_j}{|\bm x|^4} = \int \mathrm{d}^2\bm x \; \int\limits_0^\infty \mathrm{d}s\; s \ee^{-s|\bm x|^2} x_ix_j = \int \mathrm{d}s \; \left(\frac{\pi}{s}\right)^{d/2} \frac{\delta_{ij}}{2}.
\end{equation}We can evaluate this integral using dimensional regularization by splitting the integral at $s=R$.   For $s<R$, we evaluate this integral in $d=2-\epsilon$ -- this corresponds to long distances and we use an IR cutoff $L_{\mathrm{DR}}$.   For $s>R$, we evaluate this integral in $d=2+\epsilon$ -- this corresponds to short distances and we use a UV cutoff $\xi_{\mathrm{DR}}$.  We find \begin{equation}
I_{2,0ij} =\frac{\pi}{2} \delta_{ij} \left[ \frac{2}{\epsilon} L_{\mathrm{DR}}^\epsilon \left(\frac{R}{\pi}\right)^{\epsilon/2} + \frac{2}{\epsilon} \xi_{\mathrm{DR}}^{-\epsilon} \left(\frac{\pi}{R}\right)^{\epsilon/2}\right] = \delta_{ij}\left[\frac{2\pi}{\epsilon} + \pi \log \frac{L_{\mathrm{DR}}}{\xi_{\mathrm{DR}}} + \mathcal{O}(\epsilon)\right].  \label{eq74}
\end{equation}Importantly we see that this final answer is independent of $R$, as it must.


The next integral we evaluate is used to compute $\int \mathrm{d}^2\bm x\; v_1^i (\bm v_1\cdot\bm v_2)$: \begin{equation}
I_{2,2i} \equiv \int \mathrm{d}^2x  \frac{(x-X_1)_i (\bm x - \bm X_1)\cdot (\bm x- \bm X_2)}{(\bm x- \bm X_1)^4 (\bm x-\bm X_2)^2}.
\end{equation}As this integral is UV divergent, we regulate this integral by continuing to $d=2+\epsilon$, employing a UV regulator $\xi_{\mathrm{DR}}$.  Using identical substitutions to before, and following an identical procedure, we obtain \begin{align}
I_{2,2i} &= \int \mathrm{d}^2y\mathrm{d}S\mathrm{d}\alpha \; \alpha S^2 (y - (1-\alpha)X_{12})_i (\bm y - (1-\alpha)\bm X_{12})\cdot (\bm y+\alpha\bm X_{12}) \exp\left[-S\bm y^2 - S\alpha(1-\alpha)|\bm X{12}|^2\right] \notag \\
&= \xi_{\mathrm{DR}}^{-\epsilon} \int \mathrm{d}S\mathrm{d}\alpha \; \alpha S^2 \left(\frac{\pi}{S}\right)^{1+\epsilon/2} X_{12i} \ee^{-S\alpha(1-\alpha)|\bm X_{12}|^2}\left(\alpha(1-\alpha)^2|\bm X_{12}|^2 - \frac{1-\alpha}{S} + \frac{2\alpha-1}{2S}\right).
\end{align}We find that the first two terms are finite as $\epsilon\rightarrow 0$, and cancel each other exactly.  The latter term evaluates to \begin{align}
I_{2,2i} &= \int\limits_0^1 \mathrm{d}\alpha \; \xi_{\mathrm{DR}}^{-\epsilon} \Gamma\left(1-\frac{\epsilon}{2}\right) \frac{(2\alpha-1)\alpha \pi^{1+\epsilon/2}}{2(\alpha(1-\alpha))^{1-\epsilon/2}}|\bm X_{12}|^\epsilon = \left(\frac{|\bm X_{12}|}{\xi_{\mathrm{DR}}}\right)^\epsilon \frac{\pi^{(3+\epsilon)/2}\Gamma(\epsilon/2)\Gamma(1-\epsilon/2)}{2^{2+\epsilon}\Gamma((3+\epsilon)/2)} \frac{X_{12i}}{|\bm X_{12}|^2} \notag \\
&=\pi \left[ \frac{1}{\epsilon} + \log\frac{|\bm X_{12}|}{\xi_{\mathrm{DR}}} - 1 + \frac{\gamma + \log \pi}{2} + \mathcal{O}(\epsilon) \right]\frac{X_{12i}}{|\bm X_{12}|^2}
\end{align}

The final integral we evaluate is used to compute $\int \mathrm{d}^2\bm x\; (\bm v_1 \cdot \bm v_2)^2$: \begin{equation}
I_{2,3} \equiv \int \mathrm{d}^2x \; \frac{[(\bm x- \bm X_1)\cdot (\bm x - \bm X_2)]^2}{(\bm x - \bm X_1)^4 (\bm x -\bm X_2)^4}.
\end{equation}Again this is UV divergent, so we regulate this as before: \begin{align}
I_{2,3} &= \xi_{\mathrm{DR}}^{-\epsilon} \int \mathrm{d}S\mathrm{d}\alpha \; \alpha(1-\alpha)S^3 e^{-S\alpha(1-\alpha)|\bm X_{12}|^2} \left(\frac{\pi}{S}\right)^{1+\epsilon/2}\left(\frac{d(d+2)}{4S^2} - \frac{d\alpha(1-\alpha)|\bm X_{12}|^2}{S} \right. \notag \\
&\left.\;\;\;\;\;\;\;+ \frac{(2\alpha-1)^2|\bm X_{12}|^2}{2S} + \alpha^2(1-\alpha)^2|\bm X_{12}|^4\right).
\end{align}Only the third term in this sum is divergent as $\epsilon\rightarrow 0$.  We find that the first two terms exactly cancel, and that the fourth term evaluates to $2\pi |\bm X_{12}|^{-2}$.   The third term evaluates to \begin{equation}
\left(\frac{|\bm X_{12}|}{\xi_{\mathrm{DR}}}\right)^\epsilon \frac{\pi^{(5+\epsilon)/2}\csc(\pi\epsilon/2)\Gamma(2-\epsilon/2)}{2^{1+\epsilon}|\bm X_{12}|^2 \Gamma(1-\epsilon/2)\Gamma((3+\epsilon)/2)} = \left[ \frac{2\pi}{\epsilon} + \pi \left(2\log\frac{|\bm X_{12}|}{\xi_{\mathrm{DR}}} + \log \pi - 3 + \gamma \right) +\mathcal{O}(\epsilon)\right]|\bm X_{12}|^{-2}
 \end{equation}Overall we find that \begin{equation}
 I_{2,3} = \frac{\pi}{|\bm X_{12}|^2} \left(\frac{2}{\epsilon} + 2\log\frac{|\bm X_{12}|}{\xi_{\mathrm{DR}}}-1+\gamma + \log \pi\right).
 \end{equation}

Finally, we relate the dimensional regularization cutoffs $L_{\mathrm{DR}}$ and $\xi_{\mathrm{DR}}$ to more physically motivated cutoffs $L$ and $\xi$.   This can be done by defining $L$ and $\xi$ with minimal subtraction, so that \begin{subequations}\begin{equation}
\log \frac{\xi_{\mathrm{DR}}}{\xi} = -\log \frac{L}{L_{\mathrm{DR}}}   =  \frac{\log \pi - 2+\gamma}{2} - \frac{1}{\epsilon}.
\end{equation}\end{subequations}Using these identifications we find Eq. (\ref{smokering}).

\subsubsection{3-Velocity and 4-Velocity Integrals}
We could not compute these integrals analytically.   Let us nonetheless discuss their properties.

Firstly, one of the 3-velocity integrals has a UV logarithmic divergence.   Again, we can regulate this with dimensional regularization.   Using an analogous method to before, defining \begin{equation}
\bm y = \bm x - \alpha_1 \bm X_1 - \alpha_2 \bm X_2 - \alpha_3 \bm X_3
\end{equation} \begin{equation}
\mathcal{K}_3 \equiv \alpha_1\alpha_2 |\bm X_{12}|^2 + \alpha_1\alpha_3 | \bm X_{13}|^2 + \alpha_2\alpha_3 |\bm X_{23}|^2 ,
\end{equation}we obtain \begin{align}
& m^4 \int \mathrm{d}^2\bm x \; ( \bm v_1 \cdot \bm v_2 )(\bm v_1\cdot\bm v_3) = \xi_{\mathrm{DR}}^{-\epsilon} \int \mathrm{d}S \delta\left(1-\sum_{i=1}^3 \alpha_i\right) \prod_{i=1}^3 \mathrm{d}\alpha_i \; \left(\frac{\pi}{S}\right)^{1+\epsilon/2} S^3 \alpha_1 \mathrm{e}^{-\mathcal{K}_3S} \left[\frac{d(d+2)}{(2S)^2}+\right. \notag \\
&\frac{(\alpha_2 \bm X_{21}+\alpha_3\bm X_{31})^2 + (\alpha_1 \bm X_{12} + \alpha_3 \bm X_{32})\cdot (\alpha_1 \bm X_{13} + \alpha_2 \bm X_{23})  }{2S} + d\frac{(\alpha_2\bm X_{21} + \alpha_3 \bm X_{31})\cdot(\alpha_1(\bm X_{12}+ \bm X_{13}) + (\alpha_2-\alpha_3)\bm X_{23})}{2S}  \notag \\
&\left. + (\alpha_2 \bm X_{21} + \alpha_3 \bm X_{31})\cdot (\alpha_1 \bm X_{12} + \alpha_3 \bm X_{32}) (\alpha_2 \bm X_{21} + \alpha_3 \bm X_{31})\cdot (\alpha_1 \bm X_{13} + \alpha_2 \bm X_{23}) \right]
\end{align}Let us now discuss the divergences of this integral.   Divergences come from divergences in factors of $1/\mathcal{K}_3$.   The most singular region of the $\alpha$-simplex is $\alpha_1\approx 1$, due to the factor of $\alpha_1$ in front of the integral.  Near this singular region, $\mathcal{K}_3$ is linear in $\alpha_2$ and $\alpha_3$.   There is a two-dimensional simplex integral, and after power counting in $\alpha_{2,3}$ only a single term is divergent: \begin{equation}
\xi_{\mathrm{DR}}^{-\epsilon}\Gamma\left(2-\frac{\epsilon}{2}\right)\frac{\pi^{1+\epsilon/2}}{2} \int  \delta\left(1-\sum_{i=1}^3 \alpha_i\right) \prod_{i=1}^3 \mathrm{d}\alpha_i \; \bm X_{12}\cdot \bm X_{13} \frac{ \alpha_1^3 }{\mathcal{K}_3^{2-\epsilon/2}} \approx \frac{\pi}{2} \frac{\bm X_{12}\cdot\bm X_{13}}{|\bm X_{13}|^2 |\bm X_{12}|^2} \log \frac{\min(|\bm X_{12}|, |\bm X_{13}|)}{\xi}.
\end{equation}To obtain this final answer in such a simple form, we have resorted to an alternative regulariaztion scheme which directly deletes the singular region from the integrand before doing the original $\bm x$ integral, which will suffice for obtaining the logarithmically divergent contribution.\footnote{This is substantially more efficient for collecting all logarithmic divergences, but is not helpful for computing subleading terms.  All logarithmic divergences may be shown to be equal to what we have asserted, consistently within this alternative regularization scheme as well.}   In general, we are able to perform the $S$ integral analytically, but not any of the integrals on the $\alpha$-simplex.

At this point, we have shown enough to collect all of the logarithmic divergences in the action.  A simple analysis reveals that they organize themselves into the form Eq. (\ref{eq41}).

The remaining integrals over various multiples of velocities, which are strictly finite, may be treated with similar tricks as we have used, by converting the $\bm x$-integral into integrals over $S$ and $\alpha$s.   In doing so, one can reduce the answer to a series of unknown functions dependent only on the magnitude of the distance between various vortices; all dependence of the integrals on $\bm X_{12}\cdot\bm X_{13}$, e.g., can be exactly found.   As we have been unable to obtain illuminating answers for the rather complicating functions that result, we will not present explicit expressions for the intermediate manipulations.

\end{appendix}

\end{document}